\documentclass[aps, a4paper, twocolumn]{revtex4}

\usepackage{fontenc}
\usepackage[latin1]{inputenc}
\usepackage{times} 
\usepackage{amsmath,amssymb,amscd} 
\usepackage{latexsym}  
\usepackage{dsfont} 
\usepackage{color} 
\usepackage{rotating} 
\usepackage{epic} 
\usepackage{xspace}  
\usepackage{epsfig}

\newcommand{\be}{\begin{eqnarray*}} 
\newcommand{\ee}{\end{eqnarray*}}   
\newcommand{\ben}{\begin{eqnarray}} 
\newcommand{\een}{\end{eqnarray}}   

\newcommand{\bra}[1]{\langle  #1 |} 

\newcommand{\ket}[1]{| #1 \rangle}  

\newcommand{\braket}[2]{\langle  #1 | #2 \rangle} 
\newcommand{\dm}[2]{\ket{#1}\bra{#2}} 
\newcommand{\bic}[2]{\begin{pmatrix}#1\\#2\end{pmatrix}} 
\newcommand{\Ref}[1]{(\ref{#1})} 
  
\begin{document} 
 \title{\bf Production of heralded pure single photons from imperfect sources using cross phase modulation}

\author{Thomas Konrad\footnote{the corresponding author:
thomas.konrad@uni-konstanz.de}$^{1,2}$}
\author{Michael Nock$^{1}$}
\author{Artur Scherer$^{1}$}
\author{J\"urgen Audretsch$^{1}$}
\affiliation{1 Fachbereich Physik, Universit\"at Konstanz, D-78457
  Konstanz, Germany\\
2 School of Pure and Applied Physics, University of KwaZulu-Natal, Durban 4000, South Africa}
\author{}
\begin{abstract}
Realistic single-photon sources do not generate single photons with
  certainty. Instead they produce statistical mixtures of photons in Fock
  states $\ket{1}$ and vacuum (noise). We describe how to eliminate the noise
  in the output of the sources by means of another noisy source or a coherent
  state and cross phase modulation (XPM). We present a scheme which announces
  the production of pure single photons and thus eliminates the
  vacuum contribution. This is done by verifying a XPM related phase shift
  with a Mach-Zehnder interferometer. \\

\end{abstract}
\maketitle
\section{Introduction}
Light interacts in general weakly with its environment. Therefore on the one hand 
photonic
states are durable and can be optimally employed as  carrier of quantum
information. On the other hand it is more difficult to process the
information encoded in the state of light. To circumvent this obstacle
elaborated schemes have been invented to emulate such interaction by means of
linear optics and conditioned photodetection. The outstanding schemes of
Knill, Laflamme and Milburn \cite{KnillLaflammeMilburn2001} as well as teleportation and projection
synthesis \cite{Pegg98, Babichev2001} are useful tools for optical quantum
information processing. Yet all these schemes require pure single
photons, i.e. Fock states $\ket{1}$, on demand which in turn impends their
realization.

In recent years a variety of implementations for single-photon sources
has been investigated. Among them are schemes based on single molecule or atom
excitation \cite{MoernerNJP04,AlleaumeNJP04,RempeNJP04}, single ions
trapped in cavities \cite{WalterNJP04}, color centers in diamonds
\cite{Weinfurter00,Grangier02}, quantum dots \cite{Yuan02,
  SantoriNJP04}  and parametric down conversion (PDC)
\cite{Fasel04,Pittman04}.
These sources differ in the wavelength and purity of the state of the emitted
photons, their repetition rate and whether they produce a photon on
demand or heralded, i.e, announced by an event. The latter is for
example the case with PDC-sources. PDC produces randomly
photon pairs and the presence of one  photon is
indicated by the detection of the other.

None of the existing single-photon  sources, however, emits a pure single 
photon at a given time with certainty. The emission of multiple photons is
negligible for most single-photon sources, cf.\ for
example~\cite{WalterNJP04}. Therefore their output in a certain mode can be
modeled by a mixture of a Fock state $\ket{1}$ and vacuum
$\ket{0}$: 
\begin{equation}\label{imperfect}
  \rho = p\ket{1}\bra{1}+(1-p)\ket{0}\bra{0}\,,
\end{equation}
where $p$ is called the \emph{efficiency} of the single-photon
source ($0<p<1$). We refer to such mixed states as noisy photons while we call
Fock states $\ket{1}$ pure photons. Good sources have efficiencies of
$p\approx 0.6$. To the best of our knowledge the highest efficiencies reached
so far are $p=0.83$ \cite{Pittman04} and $p=0.86$ \cite{MoernerNJP04}. The
challenge is to construct a set-up by which the efficiency $p$ can be
increased in the ideal case up to one.

Investigations so far indicate that the efficiency may not be improved by means
of linear optics without adding multi-photon components. It has also been shown
~\cite{BerryScheelSandersKnight2004,BerryScheelSandersKnightLaflamme2004}
that the enhancement of $p$ is limited under these circumstances. The
efficiency $p$ cannot reach 1 given a finite
number of imperfect sources by means of linear optics and photodetection. In
addition, three or less noisy photons are not enough to obtain an improvement
at all. Employing homodyne detection and one noisy photon is not sufficient
either \cite{Berry05}. At least an enhancement of $p$
has been achieved in \cite{BerryScheelSandersKnightLaflamme2004} but at the
expense of adding a two-photon component.  

In this article we present how to purify and herald a noisy photon as given in 
(\ref{imperfect})
by means of nonlinear optics. The scheme we employ has formerly been used by
Milburn~\cite{M89}, Imoto~\cite{IHY85} and others for alternative purposes and in
different manners. It consists of a medium in which a signal and a
probe mode experience cross phase modulation. The resulting phase shift of the
probe mode is verified by means of a Mach-Zehnder interferometer. Similar
features like the use of XPM and coherent states can also be found in
\cite{Munro05}. There a QND measurement of the photon number is
proposed. However, the actual scheme can distinguish between one photon and
vacuum only up to a small non-vanishing error probability. In \cite{Konrad05}
we have proposed a method based on two-photon absorption which grants
vanishing error probability. Thus it enables the generation of a pure
photon. The scheme proposed in this article also possesses this property. The
advantage of the present scheme proposed in this article is that it allows in
principle to detect and announce a single photon emitted by the source with
arbitrarily high probability. 

This article is organized as follows. 
First we discuss cross phase modulation (Sec.~\ref{xpm}) and explain the functional principle
of our set-up (Sec.~\ref{det}). In Sec.~\ref{noisy} we study the
case in which two noisy photons are used as inputs. Then we turn to the more
realistic case in which we use one noisy  photon in the signal mode
and a coherent state in the probe mode (Sec. \ref{coherent}). Eventually, 
in Sec.~\ref{real} we discuss a potential realization 
for large cross phase modulation, which is desirable 
in our scheme. An appendix contains transparency conditions for a Mach-Zehnder
interferometer and extended schemes to generate single photons.

\section{Cross Phase Modulation} \label{xpm}

Cross phase modulation (XPM), which is also referred to as
cross Kerr interaction, is an interaction between two modes $A$ and
$B$ of a light field governed by the Hamiltonian 
\begin{equation}
H_{\mbox{\tiny XPM}}=-\chi a^\dag ab^\dag b\;,
\end{equation}
cf.~\cite{IHY85}. Here, 
$\chi$ is a real constant which is related to the third-order nonlinear 
susceptibility coefficient usually denoted by $\chi^{(3)}$, 
and $a$, $b$ represent the annihilation operators
of a photon in mode $A$ and mode $B$, respectively. 

\begin{figure}[h!]
\begin{center}
\epsfig{file=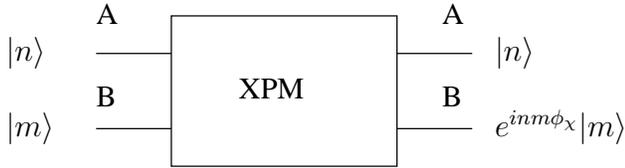}
\caption{Schematic representation of cross phase modulation between two modes 
$A$ and $B$. The action of XPM is described in terms of 
the corresponding transformation of Fock states in the 
Schr\"odinger picture.}
\label{fig1}
\end{center}
\end{figure}

Two light modes which undergo a cross phase modulation during time
$\Delta t$ aquire a phase shift which depends on the product of the 
number of photons the two modes
contain. In the Schr\"odinger picture this effect shows up e.g.\ in the
evolution of a Fock state
with $n$ photons in Mode $A$ and $m$ photons in Mode $B$:
\begin{eqnarray}
\ket{n^A,m^B}&\xrightarrow{\textrm{XPM}} 
&\exp(i\chi\Delta t a^\dag ab^\dag b)
\ket{n^A,m^B}\nonumber
\\  &=&\exp(i\phi_\chi n m)\ket{n^A,m^B} \label{XPMACTION}
\end{eqnarray}
with $\phi_\chi:=\chi \Delta t$. We exclude the case of a non-working XPM
($\phi_\chi= 2k\pi, k\in \mathbb{N}$).
Eq.~(\ref{XPMACTION}) can be interpreted as if light in mode $B$
(probe mode) experiences a phase shift due to XPM depending on 
the number of photons entering in mode $A$ (signal mode),
cf. Fig. \ref{fig1}. 

On the other hand no phase shift occurs, 
if mode $A$ contains zero photons. We employ this effect in the following to
detect and announce a single  photon in mode $A$ without absorbing it. This is
done by verifying a phase shift of light in mode $B$ by means of a
Mach-Zehnder interferometer followed by a photodetector. 

XPM naturally occurs in non-linear media (Kerr media) where the index of refraction
depends on the intensity of incoming light. Since we require vanishing
absorption rates in order not to loose incoming  photons, we are confronted
mostly with media  which  also possess very low cross phase modulation
rates ($\chi\ll 1$). This problem can be tackled by arranging long interaction
times either via electromagnetically induced transparency (EIT) or due to 
co-propagation of two modes over long distances in optical fibers. We
postpone the discussion of possible realizations of XPM to Sec. \ref{real}.      

\section{Detection of XPM-phase shift by means of a Mach-Zehnder Interferometer}
\label{det}
\begin{figure}[h!]
\begin{center}
\epsfig{file=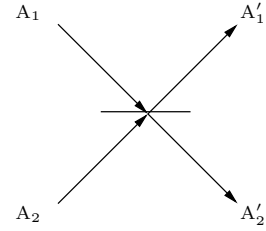}
\caption{Beam splitter with two input modes $A_1$ and
$A_2$ and two output modes $A'_1$ and $A'_2$.}
\label{fig2}
\end{center}
\end{figure}

Before we explain the functioning principle   of our scheme let us briefly
introduce our beam splitter convention. A general beam splitter with
two input modes
$A_1$ and $A_2$ and two output modes $A'_1$ and $A'_2$ is depicted in
Fig.~\ref{fig2}. 

Like in the case of cross phase modulation we represent the action of such a
beam splitter on the electromagnetic field in the Schr\"odinger picture
(cf. \cite{ScheelKnight2003}). A pure
input state given by $f(a^{\dagger}_1,a^{\dagger}_2)\ket{0,0}$ is transformed
due to the beam splitter into the output state
$f(\tilde{a}^{\dagger}_1,\tilde{a}^{\dagger}_2)\ket{0,0}$. 

Thereby the creation operators $a^{\dagger}_1$ and $a^{\dagger}_2$
corresponding to the input modes $A_1$ and $A_2$ are replaced as arguments of
the unmodified function $f$ by $\tilde{a}^{\dagger}_1$ and
$\tilde{a}^{\dagger}_2$  according to: 
\begin{eqnarray}
\label{beamsplit}
a_1^\dag &\xrightarrow{\textrm{BS}}
&\tilde{a}^\dag_1= \cos(\theta)a'^\dag_1 +
e^{-i\phi}\sin(\theta)a'^\dag_2
\nonumber\\
a_2^\dag &\xrightarrow{\textrm{BS}} &\tilde{a}^\dag_2=
-e^{i\phi}\sin(\theta)a'^\dag_1+\cos(\theta)a'^\dag_2\;,
\label{GeneralBeamsplitterTrafo}
\end{eqnarray}
cf.~\cite{ScheelKnight2003}. 
Here $a'^{\dagger}_1$, $a'^{\dagger}_2$ are the creation operators of 
the output field modes  $A'_1$ and $A'_2$, respectively. $\phi$ represents a
relative phase shift, $\cos^2(\theta)$ and $\sin^2(\theta)$ are the
reflectivity and transmittivity of the beam splitter. For the sake of simplicity, 
however, in what follows we will omit the prime labels and denote
the output field modes and the corresponding operators by the same letters as
the input modes. 

\begin{figure}[b!]
\begin{center}
\epsfig{file=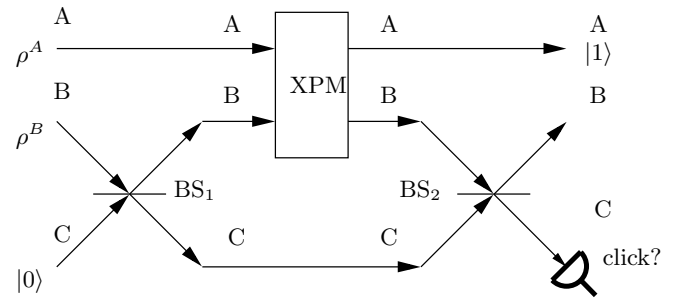}
\caption{Mach-Zehnder interferometer with cross phase modulation in the upper
  arm. A click of detector in mode $C$ announces a pure single photon Fock
  state in mode A.}
\label{fig3}
\end{center}
\end{figure}

Our scheme  to detect the production of a single photon is depicted in
Fig.\ref{fig3}. An imperfect single photon source emits in mode $A$ either a
single-photon state $\ket{1}$ with probability $p_A$ or
vacuum $\ket{0}$ with probability $1-p_A$. The resulting
statistical mixture is denoted by  
\begin{equation}
  \rho^A= p_A\dm{1^A}{1^A}+(1-p_A)\dm{0^A}{0^A}\,.
\end{equation}
$\rho^A$ enters the set-up in mode $A$. 
The input states of the auxiliary modes $B$ and $C$ of the 
Mach-Zehnder interferometer, which is composed of beam splitters $BS_1$ and $BS_2$, 
are given by $\rho^B$ and the vacuum $\ket{0^C}$, respectively. In the upper arm
of the interferometer a medium is placed which exerts a cross phase modulation
(XPM) between mode $A$ and mode $B$. 

In case mode $A$ contains only vacuum, the light in the upper arm of the
interferometer (mode $B$)  does not experience a phase shift relative to the
light in the lower arm (mode $C$). It is important that the beam splitters of
the Mach-Zehnder interferometer are adjusted such, that in this case no
photons leave the interferometer in mode $C$. Therefore the detector placed in
output $C$ cannot respond if mode $A$ does not contain a photon. On the other
hand, if the single-photon source emits a photon in mode $A$, the phase of
the light in mode $B$  is shifted due to XPM relative to the phase in Mode
$C$. This results in a non-vanishing probability for a click of the detector
in output $C$. Hence, the detection of light in mode $C$ implies the presence of a
photon in mode $A$. A selection according to the clicks of the detector yields
the preparation of a single-photon Fock state $\ket{1}$ in mode $A$. Therefore single
photons are heralded by the response of the photodetector in mode $C$.

We will address two main issues: \\
(i) The condition that a detection event in mode $C$ announces {\em with
  certainty} a pure single photon in mode $A$. \\
(ii) Provided that condition (i) is fulfilled, what is the probability to
detect a photon present in the signal mode $A$, i.e., what is the detection
efficiency? \\
The latter depends on the strength of the XPM and on the state $ \rho^B$ of
the auxiliary mode $B$. We will discuss two possible inputs in $B$: a second
noisy photon and a coherent state of light.

\section{Heralding pure single photons by means of noisy photons} \label{noisy}

Let us discuss the case that both states $\rho^A$
and $\rho^B$ which enter the set-up are noisy photons:
\begin{eqnarray}
\rho^A&=&p_A\dm{1^A}{1^A}+(1-p_A)\dm{0^A}{0^A}\,, \\
\rho^B&=&p_B\dm{1^B}{1^B}+(1-p_B)\dm{0^B}{0^B}\,. 
\end{eqnarray}  
The initial state of the signal mode $A$ and the probe mode $B$ 
thus reads 
\begin{eqnarray}
\rho^A\otimes\rho^B=&& p_Ap_B\dm{1^A1^B}{1^A1^B} \\
&&+p_A(1-p_B)\dm{1^A0^B}{1^A0^B}\nonumber\\
&& + (1-p_A)p_B\dm{0^A1^B}{0^A1^B}\nonumber\\
&&+(1-p_A)(1-p_B)\dm{0^A0^B}{0^A0^B}\,.\nonumber
\end{eqnarray}

With respect to both issues (i) and (ii) above we only have to consider the
case that a photon enters in probe mode $B$, which happens with probability
$p_B$. Otherwise, if mode $B$ contains only vacuum, no photon is present in
the Mach-Zehnder interferometer and thus no heralding of a photon in mode $A$
is possible. 

We turn to condition (i). In order to herald pure single photons in the signal
mode the detector must not respond if the signal mode contains vacuum. Whether this
condition is fulfilled can be checked by propagating an input state which
contains vacuum in the signal mode, one photon in the probe mode and 
no photon in mode $C$. Its evolution due to the Mach-Zehnder interferometer
and XPM is as a consequence of transformations ~(\ref{XPMACTION}) and
(\ref{GeneralBeamsplitterTrafo}) 
of the form:
\begin{eqnarray}
\ket{0^A1^B0^C}\xrightarrow{\textrm{BS$_2$ $\circ$ XPM $\circ$
    BS$_1$}} c_{010}\ket{0^A1^B0^C}+c_{001}\ket{0^A0^B1^C}\;.
\nonumber\\
\end{eqnarray}  
The detector in mode $C$ cannot respond if the amplitude $c_{001}$ vanishes,
i.e.,
\begin{equation}
c_{001}=e^{-i\phi_2}\cos(\theta_1)\sin(\theta_2)+ 
e^{-i\phi_1}\sin(\theta_1)\cos(\theta_2)=0\,.
\end{equation}
Here $\cos^2(\theta_{1,2})$ and $\sin^2(\theta_{1,2})$ are the
reflectivity and transmittivity of beam splitters $BS_1$ and $BS_2$,
respectively (cf.\ Eq.(\ref{GeneralBeamsplitterTrafo})). This leads to one of
the following two constraints:
\begin{small}
 \begin{eqnarray}
\phi_1-\phi_2=2k\pi \;\;\;\quad\quad&\mbox{and}&\;  \theta_1+\theta_2=l \pi, \; k,l \in
 \mathbb{Z}\,,\label{condA} \\
\phi_1-\phi_2=(2k+1)\pi \;&\mbox{and}&\;  \theta_1-\theta_2=l \pi, \; k,l \in
 \mathbb{Z}\,. \label{condB}
\end{eqnarray}
\end{small}
It is shown in the Appendix \ref{appendixA} that if one of these constraints
 is satisfied, any state $\ket{0^A}$ $\ket{\psi}^{BC}$ entering the
 interferometer does not change.

We choose beam splitters BS$_1$ and BS$_2$ according to either constraint
\Ref{condA} or \Ref{condB} and assume that the detector has no dark
counts. Then the conditional probability for a click given that the 
signal mode $A$ contains just vacuum is zero, i.e., 
$p(\textrm{click}|0_\textrm{in}^A)=0$, which is equivalent to saying
that, provided  a click occurs there must be a photon in mode $A$. 
Hence, the conditional probability to find one photon outgoing 
in mode $A$ if the detector clicks is equal to one:
\begin{equation}
p(1_\textrm{out}^A|\textrm{click})=1\;.
\end{equation}
Thus condition (i) is fulfilled. We have obtained the following
result: Selecting the cases in which the detector clicks 
amounts to the preparation of the pure one-photon state $\ket{1^A}$ in
mode $A$. This is independent of the parameters $\phi_\chi$ of the XPM and the
efficiency $p_B$ of $\rho^B$, respectively.

However, with regard to the practical use of the set-up the question of issue
(ii) remains, namely, how probable it is that a photon in the signal mode $ A$ is detected, i.e. causes a
click of the detector. This depends on the strength $\phi_\chi$ of the XPM and on the
efficiency $p_B$ of the source feeding probe mode $B$. 

The action of the Mach-Zehnder interferometer combined with the XPM medium on
the input state $\ket{1^A1^B0^C}$ is given by the transformation (cf.\ Eq.~(\ref{XPMACTION}) and
(\ref{GeneralBeamsplitterTrafo})):
\begin{equation}
\ket{1^A1^B0^C}\xrightarrow{\textrm{BS$_2$ $\circ$ XPM $\circ$
      BS$_1$}}c_{110}\ket{1^A1^B0^C}+
c_{101}\ket{1^A0^B1^C}
\end{equation}
with
\begin{eqnarray}
c_{110}&=&
e^{i\phi_\chi}\cos(\theta_1)\cos(\theta_2)-e^{-i(\phi_1-\phi_2)}\sin(\theta_1)\sin(\theta_2)\,,\nonumber\\ 
c_{101}
&=& e^{i(\phi_\chi-\phi_2)}\cos(\theta_1)\sin(\theta_2)+e^{-i\phi_1}\sin(\theta_1)\sin(\theta_2)\;.\nonumber\\&&
\end{eqnarray}
Inserting either constraint \Ref{condA} or \Ref{condB} leads to the probability
for a click of the detector given by:
\begin{equation}
  p(\mbox{click}|1^A_\textrm{in}1^B_\textrm{in}0^C_\textrm{in})
=|c_{101}|^2=\sin^2(\tfrac{\phi_\chi}{2})\sin^2(2\theta_1)\,.
\end{equation}
Its maximal value
\begin{equation}
p(\mbox{click}|1^A_\textrm{in}1^B_\textrm{in}0^C_\textrm{in})=\sin^2(\tfrac{\phi_\chi}{2})
\end{equation}
is thus achieved for $\theta_1=\frac{\pi}{4}$ or $\theta_1=\frac{3\pi}{4}$. In both
cases we are free to choose $\phi_1-\phi_2=0$ or $\phi_1-\phi_2=\pi$. The
corresponding $\theta_2$ follows from Eq. \Ref{condA} or \Ref{condB},
respectively. In all these cases beam splitters BS$_1$ and BS$_2$ are
symmetric. Please note, that the optimal choice of beam splitters does not
depend on the exerted phase shift $\phi_\chi$. Since we assume $\phi_\chi\neq
2k\pi$ (with $k \in \mathbb{N}$), the detector responds in some of the cases
when a single photon enters in the signal mode, while it does not so if this
mode is empty. We have thus obtained a heralded single-photon source with
efficiency $p=1$.

The probability $P_E$ that a photon present in the signal mode is successfully
detected and announced by a click of the detector in output $C$ (i.e., the
detection efficiency, cf. question (ii)) amounts to 
\begin{equation} \label{pe18}
  P_E=p(\mbox{click}|1^A_\textrm{in}1^B_\textrm{in}0^C_\textrm{in})p_B=\sin^2(\tfrac{\phi_\chi}{2})p_B,
\end{equation}
where $p_B$ is the probability that one photon enters in mode $B$. 
$P_E$ is the detection efficiency. This leads to the total probability $P_T$ to produce a
heralded single photon from the output of two imperfect single photon
sources. It is given by $P_T=P_Ep_A$, where $p_A$ is the efficiency of the source
feeding mode $A$. 

Although it is in principle possible to generate a heralded
pure photon from two imperfect sources using our scheme, the detection
efficiency $P_E\propto \sin^2(\tfrac{\phi_\chi}{2})$ is low if the phase shift
$\phi_\chi$ corresponding to the cross phase modulation is small (cf. also
Sec. \ref{real}). This
disadvantage can be partly compensated by using intensive laser light instead
of a noisy photon as input of mode $B$.  
 
\section{Heralding pure single photons by means of coherent states}
\label{coherent}

In this section we explore the possibility to produce a heralded pure photon
from a mixture $\rho^A=p_A\dm{1^A}{1^A}~{+(1-p_A)\dm{0^A}{0^A}}$ with 
the set-up described in Sec. \ref{det} and a coherent
state $\rho^B=\dm{\beta}{\beta}$ as input of mode $B$. For this purpose we
have to ensure that the detector in mode $C$ cannot click if mode $A$ contains
just vacuum. This leads to the same constraints (\ref{condA}) and
(\ref{condB}) for the beam splitters $BS_1$ and $BS_2$, which we obtained in
the previous section (cf. the transparency conditions for a Mach-Zehnder
interferometer in Appendix \ref{appendixA}).

Hence, provided  beam splitters $BS_1$ and $BS_2$ are chosen according to
either Eq.~(\ref{condA}) or Eq.~(\ref{condB}) and the detector has no dark counts,
the conditional probability to find one photon outgoing in mode $A$ if the
detector clicks is again one
\begin{equation} \label{ptuer}
p(1_\textrm{out}^A|\textrm{click})=1\;.
\end{equation}
As in the previous section \ref{noisy} selection according to the clicks of
the detector amounts to the preparation of the pure single-photon Fock state
$\ket{1^A}$ (cf. issue (i)). But what is the probability of such a
preparation? We now calculate the detection efficiency (cf. issue (ii)).

If a photon is present in the signal mode $A$ we obtain the following
state transition
{\small
\begin{eqnarray}
&&\ket{1}^A\ket{\beta}^B\ket{0}^C\nonumber\\
\xrightarrow{BS_1}&&\ket{1}^A\ket{\beta\cos(\theta_1)}^B\ket{\beta
  e^{-i\phi_1}\sin(\theta_1)}^C\nonumber\\
\xrightarrow{XPM}&&
\ket{1}^A\ket{e^{i\phi_\chi}\beta\cos(\theta_1)}^B\ket{\beta e^{-i\phi_1}\sin(\theta_1)}^C\nonumber\\
\xrightarrow{BS_2}&& \ket{1}^A\ket{e^{i\phi_\chi}\beta\cos(\theta_1)\cos(\theta_2)-\beta
  e^{i(\phi_2-\phi_1)}\sin(\theta_1)\sin(\theta_2)}^B \nonumber\\&&\ket{\beta
e^{i(\phi_\chi-\phi_2)}\cos(\theta_1)\sin(\theta_2)+\beta e^{-i\phi_\chi}\sin(\theta_1)\cos(\theta_2)}^C\nonumber\\
&&=\ket{1}^A\underbrace{\ket{\beta(e^{i\phi_\chi}\cos^2(\theta_1)+\sin^2(\theta_1))}^B}_{=:\ket{\beta'}^B}\nonumber\\&&\quad\quad\otimes
\underbrace{\ket{\beta e^{-i\phi_2}\tfrac12\sin(2\theta_1)(1-e^{i\phi_\chi})}^C}_{=:\ket{\gamma'}^C}\,.
\end{eqnarray}
}
In order to obtain the last line we have inserted constraint
\Ref{condA}. The outgoing state is separable being a product of
coherent states in modes $B$ and $C$ as expected for classical fields which
pass through an interferometer. 

Based on this outgoing state we  calculate the
probability $p(\mbox{click}|1^A_\textrm{in}\beta^B_\textrm{in}0^C_\textrm{in})$ 
for a response of the detector given that one photon entered the setup in mode
$A$. We assume here that the response of the detector corresponds to the effect operator
$\hat{P}_{\textrm{C,det.}}=\sum_{k=1}^\infty \dm{k^C}{k^C}$. Thus the related 
probability amounts to:  
\begin{eqnarray}
p(\mbox{click}|1^A_\textrm{in}\beta^B_\textrm{in}0^C_\textrm{in})&=&
{}^C\braket{\gamma'}{\hat{P}_{\textrm{C,det.}}|\gamma'}^C\nonumber \\
&=&1-|{}^C\braket{0}{\gamma'}^C|^2\nonumber \\
&=& 1-e^{-|\beta|^2\sin^2(2\theta_1)\sin^2(\frac{\phi_\chi}{2})}\:.
\end{eqnarray}
This is the detection efficiency $P_E$ to successfully detect a photon present
in the signal mode. Constraint \Ref{condB} leads to the same expression for
$P_E$. It assumes its maximal value for the same choice of beam splitters as
in Sec. \ref{noisy}: 
\begin{equation}
P_E=p(\mbox{click}|1^A_\textrm{in}\beta^B_\textrm{in}0^C_\textrm{in})= 
1-e^{-|\beta|^2\sin^2(\frac{\phi_\chi}{2})}\,.
\end{equation}
This result is to be compared with $P_E$ of Eq. \Ref{pe18}.
Since the detector does not click if no photon is present in the signal mode,
but does respond with a finite probability $P_E$ (provided that $\phi_\chi
\neq 2k\pi$) if the signal mode contains a single photon, it is possible to
generate pure heralded photons with our set-up using a coherent state in mode
$B$. The resulting single-photon source has the efficiency $p=1$.
The total probability $P_T$  to obtain a heralded pure photon from a source
with efficiency $p_A$ then amounts to $P_T=P_Ep_A$.   

\begin{figure}
\begin{center}
\epsfig{file=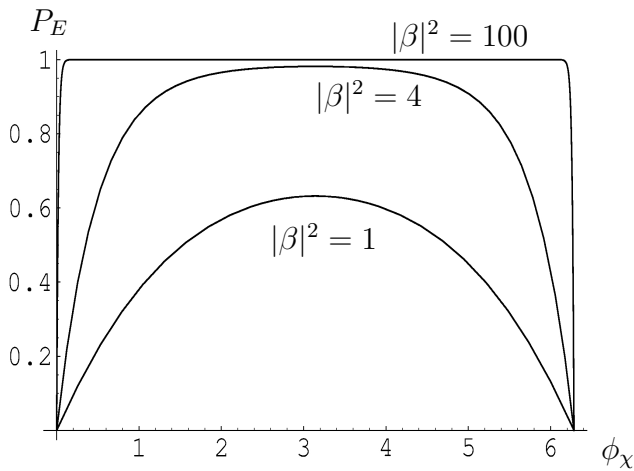}
\caption{Detection efficiency $P_E$ as function of the phase shift
  $\phi_\chi$ due to XPM for three different values of $|\beta|$.}
\label{fig4}
\end{center}
\end{figure}

The ability to produce pure photons from an imperfect source heralded
by a photodetection distinguishes our scheme among others. Its quality depends
on the probability $P_E$ to announce a single photon present in the signal
mode. It crucially depends on the product $|\beta|^2\sin^2(\frac{\phi_\chi}{2})$. 
For any phase shift $\phi_\chi$ the detection efficiency $P_E$ can be
increased arbitrarily close to $1$ by choosing a sufficiently high mean photon
number $|\beta|^2$ (cf. Fig. \ref{fig4}). It can be seen that $P_E$ increases
rapidly already for small values of $|\beta|^2$. 

\section{Realization of XPM}
\label{real} 
\begin{figure}[h!]
\begin{center}
\epsfig{file=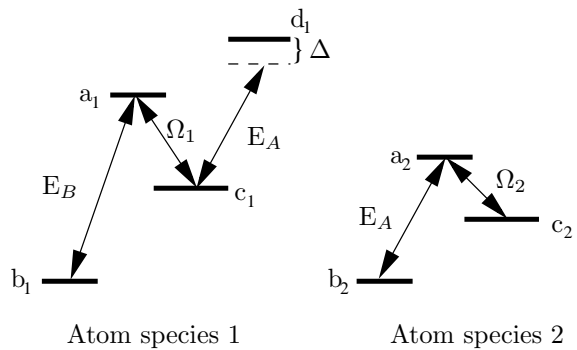}
\caption{Atomic level configuration for establishing 
double electromagnetically induced transparency and large 
XPM between the fields $E_A$ and $E_B$ of modes A and B 
within a Kerr medium (cf.~\cite{LukinImamoglu2000}). 
The two fields  $E_B$, $E_A$ 
and two classical fields $\Omega_1$, $\Omega_2$, 
are to be in resonance with $b_i\leftrightarrow a_i$ 
and  $c_i\leftrightarrow a_i$ transitions  ($i=1,2$) 
within atoms of species 1 and 2 as depicted. Quantum interference 
induced by the classical driving fields $\Omega_1$, $\Omega_2$, 
entails sharp transmission resonances of the fields 
 $E_A$, $E_B$. In this way both  $E_A$ and $E_B$ experience 
EIT. The field $E_A$ couples non-resonantly another optically 
allowed transition  $c_1\leftrightarrow d_1$ with a 
detuning $\Delta$ within the atoms 
of species 1. The induced Stark shift of level $c_1$ 
is responsible for the modification 
of the refractive index of field  $E_B$, resulting in XPM
between the fields $E_A$ and $E_B$.}
\label{fig5}
\end{center}
\end{figure}

The objective of every possible realization scheme of our proposal 
is to produce giant Kerr nonlinearities so as to make XPM as large 
as possible, even for field intensities corresponding to that of a 
single photon. In the ideal case we would like to be able to 
choose phase shifts  $\phi_\chi$ on the order of $\pi$. 
In the following we report on a promising
XPM scheme~\cite{SchmidtImamoglu1996,LukinImamoglu2000} 
that makes such huge phase shifts feasible, even if light pulses 
of ultra-small energies, i.e.\ single photons, are involved. 
It is based on the electromagnetically induced transparency (EIT) 
phenomenon~\cite{Harris1997} which makes possible to resonantly
enhance the Kerr nonlinearity $\chi^{(3)}$  
along with simultaneous elimination of absorption losses 
due to vanishing linear susceptibilities~\cite{SchmidtImamoglu1996}. 
The principle of the method is illustrated in Fig.~\ref{fig5}. 

The Kerr medium consists of two atomic species 1 and 2 
which resonantly interact with the propagating fields  
$E_A$ and $E_B$ of modes A and B as depicted in 
Fig.~\ref{fig5}.  $E_B$ is tuned to resonance 
with the atomic transition $b_1\leftrightarrow a_1$  of 
species 1 whereas $E_A$ with the  atomic transition 
$b_2\leftrightarrow a_2$  of species 2. Both atomic species 
are in addition resonantly driven by strong classical 
fields $\Omega_1$ and $\Omega_2$, which couple 
the atomic transitions   $c_1\leftrightarrow a_1$ and 
$c_2\leftrightarrow a_2$, respectively. The quantum interferences 
created by the classical driving fields involve sharp 
transmission resonances for the corresponding quantum fields 
$E_A$ and $E_B$. By this means EIT is established for both fields. 
We get double EIT (DEIT). As a consequence, both $E_A$ and $E_B$ 
propagate without absorption losses or refraction, and their group 
velocities are considerably reduced.  Large Kerr nonlinearities 
leading to cross phase modulation between the fields $E_A$ and
$E_B$ are obtained via Stark effect. The signal field $E_A$ is 
non-resonantly coupled to another optically allowed transition  
$c_1\leftrightarrow d_1$ with a detuning $\Delta$ within the atoms 
of species 1. This results in a Stark shift of level $c_1$, 
thus involving a change of the refractive index of field $E_B$. 
Since the  refractive index dispersion is very strong near 
resonances, relatively small Stark shifts are sufficient to 
induce a large index change.  The Kerr nonlinearities accomplished 
in this way have been shown to yield $\chi^{(3)}$-values that are 
orders of magnitude higher than in conventional 
systems~\cite{SchmidtImamoglu1996}. Moreover, the resulting 
XPM of the fields $E_A$ and $E_B$ can be sustained for a 
very long interaction time. As already mentioned, due to DEIT 
both fields propagate without absorption losses and with 
strongly reduced group velocities. Furthermore, the experimental 
conditions can be arranged such that their group velocities become 
equal~\cite{LukinImamoglu2000}. This in turn involves a potentially very 
long interaction time between the two fields $E_A$ and $E_B$ 
thus making possible very large conditional nonlinear phase shifts. 
As shown by Lukin and Imamo$\check{\mbox{g}}$lu in
~\cite{LukinImamoglu2000} this realization scheme 
makes feasible XPM phase shifts  of the order 
of $\pi$, even if single-photon fields are
involved.  The requirement to be fulfilled is 
$\tau_g\Delta \omega_{\mbox{\tiny max}}\gg 1$. 
Here $\tau_g$ is the group delay and 
$\Delta \omega_{\mbox{\tiny max}}$ the  
bandwidth of the EIT resonance. More details can be found 
in~\cite{SchmidtImamoglu1996,LukinImamoglu2000}. 
Finally, it has recently been shown~\cite{WangMarzlinSanders2006} 
that DEIT and large XPM between slowly co-propagating weak fields  
may also be obtained using only one atomic species. Unfortunately, however,
the XPM phase shift between two photons as estimated by the authors 
of \cite{WangMarzlinSanders2006} is considerably smaller 
than the estimates obtained by Lukin and  Imamo$\check{\mbox{g}}$lu
in~\cite{LukinImamoglu2000}. 

An experimental implementation of our scheme by means of DEIT 
would require an analysis that allows for unavoidable losses 
due to spontaneous emission processes as well as pump field
fluctuations.  These losses being small, they nevertheless 
might limit to some extent the ability of the proposed scheme 
to improve the efficiency of single photon sources. A full 
analysis of DEIT that takes into account all the various 
loss mechanisms including spontaneous emission noise 
and pump field fluctuations is beyond the scope of 
the present paper. 
Moreover, to the best of our  knowledge, 
such a comprehensive analysis does not exist in 
the literature. 
Here we take into account losses phenomenologically, 
in terms of a finite probability of absorption 
of the single-photon and a corresponding attenuation 
of the amplitude of the coherent state, respectively.   
Absorption losses are mainly caused by spontaneous 
decay of the excited states. Another source of 
absorption is introduced by the decay of coherence 
between the ground-state levels, i.e.\ decoherence 
of the dark state. 
We examine the extent of negative impact these absorption 
losses entail on our scheme. 
In particular, 
we estimate  an upper bound on the amount of spontaneous emission 
noise that can be tolerated, so as our scheme still works. 
The heuristic analysis below refers to the 
purification scheme that uses coherent states, 
as discussed in Sec.~\ref{coherent}.

Without expanding on the various loss mechanisms 
let us take into consideration losses in terms of the 
following sensible heuristic assumption. We assume 
that there is a finite probability  $\wp_a$ that a 
photon is absorbed. In particular, 
we make the  following heuristic ansatz 
which is reasonable with regard to the light field 
evolution equations suggested in \cite{LukinImamoglu2000}:
\begin{equation}\label{Eq:absorption}
\langle\hat{n}\rangle_{\mbox{\tiny out}}=
(1-\wp_a)\langle\hat{n}\rangle_{\mbox{\tiny in}}\;.
\end{equation} 
In the above equation, $\langle\hat{n}\rangle_{\mbox{\tiny in}}$ 
is the initial mean photon number of the light 
pulse when it enters the medium, while  
$\langle\hat{n}\rangle_{\mbox{\tiny out}}$ denotes the mean photon number 
of the outgoing field. The mean photon number decreases 
when the light pulses propagate through the Kerr 
medium. It is assumed to be attenuated by the factor $(1-\wp_a)$, where 
$\wp_a$ is a probability $(0\le \wp_a\le 1)$. 
In order to allow for absorption losses according 
to  Eq.~(\ref{Eq:absorption}) we modify the usual 
XPM state transformation by the following 
heuristic equations: 
{\small 
\begin{eqnarray}
\ket{1}^A\ket{\beta'}^B 
&\xrightarrow{\mbox{XPM}}& 
\sqrt{(1-\wp_a)}
\ket{1}^A\ket{\beta'\sqrt{1-\wp_a}e^{i\phi_\chi}}^B\times  \mbox{h.c.}\nonumber\\
&&+
\sqrt{\wp_a}\ket{0}^A\ket{\beta'\sqrt{1-\wp_a}}^B\times \mbox{h.c.}\nonumber\\
\ket{0}^A\ket{\beta'}^B&\xrightarrow{\mbox{XPM}}& 
\ket{0}^A\ket{\beta'\sqrt{1-\wp_a}}^B\;.
\end{eqnarray}
}
These state transformations comprise both 
the usual XPM interaction leading to a 
phase shift of the coherent state in mode $B$  
if a photon is present in mode $A$ and the 
required absorption losses in mode $A$ as well 
as in mode $B$. The absorption of photons from 
the coherent state in mode $B$ is described by 
the attenuation $|\beta'|\rightarrow\sqrt{1-\wp_a} |\beta'|$, 
which is reasonable because of 
$\langle\hat{n}^B\rangle_{\mbox{\tiny in}}=|\beta'|^2$.
Please note that there is an 
attenuation of the coherent state in mode $B$ even if 
there is no photon in mode $A$. In the above model we 
have made the approximation that either a full phase 
shift $e^{i\phi_\chi}$ is acquired or no phase 
shift at all.

The existence of absorption losses involves an imperfect 
operation of our scheme. The mere possibility of absorption 
leads eventually to faulty clicks of the detector, i.e. 
detector clicks even if there is no photon 
leaving the setup in mode $A$. As a consequence,  
the efficiency $p'_A$ of the resulting heralded 
single-photon source becomes less than one.  
Yet, we have examined the conditions under which 
an improvement of the efficiency, 
i.e.\ $p'_A>p_A$, is still possible. 
Using the above heuristic assumptions and methods of 
Sections~\ref{det} and \ref{coherent} we have  
calculated an upper bound on $\wp_a$ up to which 
absorption losses can be tolerated, so as our 
scheme to improve the efficiency of single photon 
sources still works. This upper bound depends on 
the XPM phase shift $\phi_\chi$ as well as on 
$|\beta|^2$, i.e.\ the mean photon number
in mode $B$. The dependence on $\phi_\chi$ 
is appreciable only for low mean photon numbers, 
it becomes less important with increasing value of 
$|\beta|^2$. Being not too optimistic, let us 
assume $\phi_\chi=10$ mrad, the phase shift which has been 
estimated by the authors of Ref.~\cite{WangMarzlinSanders2006}.  
Using this value the following upper bounds on $\wp_a$ have 
been obtained,  depending on the mean photon number of the 
coherent state:  
$\wp^{\mbox{\tiny max}}_a\approx 0.008$ in case 
$|\beta|^2= 10^6$, $\wp^{\mbox{\tiny max}}_a\approx 0.020$ 
in case $|\beta|^2=10^4$ and $\wp^{\mbox{\tiny max}}_a\approx 0.021$ 
in case $|\beta|^2=10^2$. Thus it appears that for $\phi_\chi=10$ mrad
 absorption losses of not more than about 2\% are acceptable. 
Larger absorption losses could be tolerated 
if low mean photon numbers are employed and higher XPM phase shift 
values were feasible. In the ideal case $\phi_\chi\approx \pi$ the 
following upper bounds have been attained with 
our heuristic model: $\wp^{\mbox{\tiny max}}_a\approx 0.06$ in case 
$|\beta|^2= 10^4$, $\wp^{\mbox{\tiny max}}_a\approx 0.35$ 
in case $|\beta|^2=10^2$ and $\wp^{\mbox{\tiny max}}_a\approx 0.80$ 
in case $|\beta|^2=1$.

We conclude that XPM schemes based on DEIT provide promising
realizations to implement the heralded single-photon generator 
via Kerr effect as proposed in this paper.  

\section{Conclusion}
We have suggested a scheme to produce  a pure single photon from the output
of an imperfect single-photon source given by the mixed state 
$p\ket{1}\bra{1}+(1-p)\ket{0}\bra{0}$ with finite efficiency $p$ which may be
arbitrarily small.
Heralded production of a single photon has been achieved in two respects. First of all we
have derived the interferometer adjustments which ensure that a detector
click indicates a single photon with certainty. Conditioning on the clicks of
the detector leads already to $p=1$. Secondly we have shown that the detection
efficiency can be made arbitrarily high.

Using noisy photons in both
signal and probe mode causes a low detection efficiency for a single 
photon in the signal mode. This problem can be overcome by enforcing 
higher interaction via DEIT or other techniques. This
disadvantage has been shown to be naturally attenuated by using a 
coherent state in the probe mode. In this case the detection
efficiency $P_E$ depends on the product
$|\beta|^2\sin^2(\frac{\phi_\chi}{2})$. 
By choosing the mean photon number $|\beta|^2$ sufficiently high $P_E$ 
can be increased arbitrarily close to one. 

\vspace*{3mm}
\noindent{\bf Acknowledgment:} We would like to thank Christian
Kasztelan, Rudolf Bratschitsch, Karl-Peter Marzlin and Alfred Leitenstorfer 
for helpful discussions. This work was supported by the \lq\lq Center
for Applied Photonics'' (CAP)  at the University of Konstanz.  

\section{Appendix}
\subsection{Transparency conditions for a Mach-Zehnder interferometer} \label{appendixA}
We consider a Mach-Zehnder interferometer which is composed of the two beam
splitters BS$_1$ and BS$_2$. Here we show how these beam splitters have to
be adjusted in order that arbitrary ingoing states of light do not change
under the interferometer's action.

Any pure state of light entering two modes $B$ and $C$ (cf. Fig.~\ref{fig3})
including an entangled state can be expressed
as a function $f(b^\dag, c^\dag)$ of the creation operators $b^\dag$ and
$c^\dag$ acting on the vacuum $\ket{0,0}$. As pointed out in
  Sec. \ref{noisy}, the action of both beam splitters in the Schrödinger
  picture can then be written as
\ben \label{trafo}
f(b^\dag, c^\dag)\ket{0,0}\xrightarrow{BS_1}f(\tilde{b}^\dag,\tilde{c}^\dag)\ket{0,0}
\xrightarrow{BS_2}f(\tilde{\tilde{b}}^\dag,\tilde{\tilde{c}}^\dag)\ket{0,0}\;. \nonumber\\
\een
Hereby the function $f$ does not change, but its operator-valued arguments are transformed
according to the transformation rules
\begin{eqnarray}\label{trafo2}
\begin{pmatrix}
\tilde{\tilde{b}}^\dag \\ \tilde{\tilde{c}}^\dag
\end{pmatrix}=U_2\begin{pmatrix}
\tilde{b}^\dag \\ \tilde{c}^\dag
\end{pmatrix}=U_2U_1\begin{pmatrix}
b^\dag \\ c^\dag
\end{pmatrix}\;.
\end{eqnarray}
$U_1$ and $U_2$ are unitary transformations which can be read
off from Eq. \Ref{beamsplit}.

Input state and output state in transformation \Ref{trafo} are equal for
arbitrary functions $f$ if $\tilde{\tilde{b}}^\dag=b^\dag$ and $\tilde{\tilde{c}}^\dag=c^\dag $. As
can be seen from Eq. \Ref{trafo2} this transparency condition is fulfilled if $U_2U_1=\openone$, or
equivalently
\ben \label{reverse}
U_2=U_1^\dag\;.
\een
This means that beam splitter BS$_2$ reverses the action of BS$_1$.
Comparing the matrix elements of $U_2$ and $U_1$,
\begin{eqnarray}
&&U_2=
\begin{pmatrix}
\cos(\theta_2) & e^{-i\phi_2}\sin(\theta_2)\\
-e^{i\phi_2}\sin(\theta_2)& \cos(\theta_2)
\end{pmatrix},\\
&&U_1^\dag=\begin{pmatrix}
\cos(\theta_1) & -e^{-i\phi_1}\sin(\theta_1)\\
e^{i\phi_1}\sin(\theta_1)& \cos(\theta_1)
\end{pmatrix}\,,
\end{eqnarray}
we obtain the following two sets of constraints
\begin{small}
\begin{eqnarray}
 \phi_1-\phi_2=2k\pi \;\;\;\quad\quad&\mbox{and}&\;  \theta_1+\theta_2=l \pi, \; k,l \in
 \mathbb{Z}\,, \\
\phi_1-\phi_2=(2k+1)\pi \;&\mbox{and}&\;  \theta_1-\theta_2=l \pi, \; k,l \in
 \mathbb{Z}\,.
\end{eqnarray}
\end{small}
Note that we already have obtained the same conditions in Sec. \Ref{det} for
a special input state. 

\subsection{Extended schemes} \label{appendixB}
Starting from our basic scheme as discussed in this article, we now construct
extended schemes.

\begin{figure}[h!]
\begin{center}
\epsfig{file=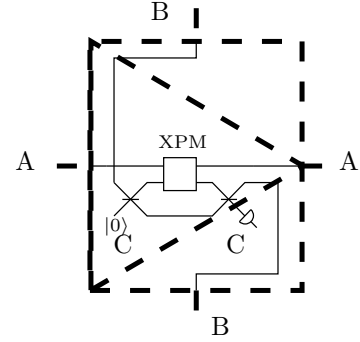}
\caption{The dashed triangle symbol is used as a replacement for the
  solid drawn single set-up.}
\label{fig6}
\end{center}
\end{figure}

The setup of Sec. \ref{det} constitutes a basic building block,
cf. Fig.~\ref{fig6}. By a suitable combination of several such basic modules,
we can increase the probability to generate a single photon. The combined
set-ups discussed below do not process the noisy photons and the coherent
state independently. Rather, they use resources more efficiently. And by
considering more noisy photons it becomes obvious how fast the probability to
purify at least one of them increases with their number.

\begin{figure}[h!]
\begin{center}
\epsfig{file=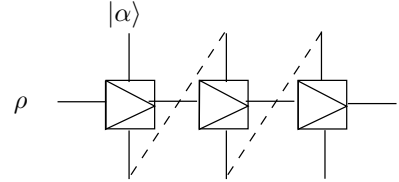}
\caption{A noisy photon and a coherent state enter the first set-up like in
  the basic scheme. Its outputs are reused as inputs of the second set-up, and
  so on.}
\label{fig7}
\end{center}
\end{figure}

The first combined scheme is depicted in Fig.~\ref{fig7}. It is intended to
purify one noisy photon with a higher
probability than a single set-up does. This is efficiently achieved by
reusing the outgoing coherent state again in a second set-up and so on. 

Let the input state be
\ben
\rho=p\dm{1}{1}+(1-p)\dm{0}{0}\;.
\een 
In the case that no photon enters from left the coherent state will
not cause a detection. But if there is a photon entering the first set-up it
will not necessarily cause a click. Instead there is still a chance that the
coherent state leaving mode C is projected onto vacuum by detection. 
Nonetheless the amplitude of the coherent state leaving Mode B would
be decreased by a factor of $\cos(\tfrac{\phi_\chi}{2})$. The probability
$p_n$ for the first click to happen in the n-th set-up is given by
\ben
p_n&=&\prod_{i=0}^{n-2}
\exp(-|\alpha|^2\sin^2(\tfrac{\phi_\chi}{2})\cos^{2i}(\tfrac{\phi_\chi}{2}))\\
&&\times\left(1-\exp\left(-|\alpha|^2\sin^2(\tfrac{\phi_\chi}{2})\cos^{2(n-1)}(\tfrac{\phi_\chi}{2})\right)\right)\;.\nonumber
\een
We obtain the probability $P_T$ for heralding one photon by summing over
  all $N$ set-ups and weighing the sum with efficiency $p$. The resulting
  probability is arbitrarily close to $p$:
\ben
P_T=p \sum_{n=1}^N p_n \xrightarrow{|\alpha|\gg 1, N \gg 1} p\;.
\een

\begin{figure}[h!]
\begin{center}
\epsfig{file=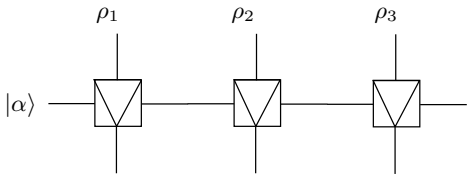}
\caption{The second combined scheme uses several noisy photons and one
  coherent state.}
\label{fig8}
\end{center}
\end{figure}

In the second combined scheme, which is illustrated in Fig.~\ref{fig8},
several single-photon sources are to be
processed with one coherent state. Similarly to the first set-up we look for
the probability that at least one photon is heralded.
The probability $p_n$ for the first click to happen in the n-th set-up can be
expressed as
\ben
p_n&=&\sum_{k=0}^{n-1}\bic{n-1}{k}p^{k}(1-p)^{n-k-1} \\&&\times
  \exp\left[-|\alpha|^2\sin^2(\tfrac{\phi_\chi}{2})
    \frac{\cos^{2k}(\tfrac{\phi_\chi}{2})-1}{\cos^2(\tfrac{\phi_\chi}{2})-1}\right]\nonumber
\\&&\times p(1-\exp\left[-|\alpha|^2
  \cos^{2k}(\tfrac{\phi_\chi}{2})\sin^2(\tfrac{\phi_\chi}{2})\right]\;.\nonumber
\een
The probability $P_T$ for heralding at least one photon is obtained
by summing over all single-photon sources. It tends to $1$ for $|\alpha|, N
\gg 1$,
\ben
P_T=\sum_{n=1}^N p_n \xrightarrow{|\alpha|\gg 1, N \gg 1} 1\;.
\een


\begin{thebibliography}{10}
\bibitem{KnillLaflammeMilburn2001}
E.\ Knill, R.\  Laflamme and G.\ J.\ Milburn, Nature {\bf 409}, 46 (2001).

\bibitem{Pegg98} D. T. Pegg, L. S. Phillips and S. M. Barnett, Phys. Rev. Lett. {\bf
    81}, 1604-1606 (1998).

\bibitem{Babichev2001} S. A. Babichev, J. Ries and A. I. Lvovsky, 
Europhys. Lett.\ {\bf 64} (1) , pp. 1-7 (2003).

\bibitem{MoernerNJP04} 
W.\ E.\ Moerner, New J.\ Phys. {\bf 6}, 88 (July 2004).

\bibitem{AlleaumeNJP04} 
R.\ All\'{e}aume, F.\ Treussart, J.\ M.\ Courty and
  J.\ F.\ Roch, New J. Phys.\ {\bf 6}, 85 (2004).

\bibitem{RempeNJP04} M.\ Hennrich, T.\ Legero, A.\ Kuhn and G.\ Rempe, 
New J.\ Phys.\ {\bf 6} (July 2004).

\bibitem{WalterNJP04}
M.\ Keller, B.\ Lange, K.\ Hayasaka, W.\ Lange and H. Walther, 
New J. Phys. {\bf 6}, 95 (July 2004). 

\bibitem{Weinfurter00} 
C.\ Kurtsiefer, S.\ Mayer, P.\ Zarda, and H.\ Weinfurter, 
Phys.\ Rev.\ Lett.\ {\bf 85}, 290-293 (2000).

\bibitem{Grangier02} 
A.\ Beveratos, S.\ Kuhn, R.\ Gacoin, J.\ P.\ Poizat, P.\ Grangier,  
Eur.\ Phys.\ J.\ D {\bf 18} 191 (2002). 

\bibitem{Yuan02} 
Z.\ Yuan, B.\ E.\ Kardynal, R.\ M.\ Stevenson, A.\ J.\ Shields, C.\
  J.\ Lobo, K.\ Cooper, N.\ S.\ Beattie, D.\ A.\ Ritchie, M.\ Pepper,
Science, Vol.\ {\bf 295}, Issue 5552, 102-105, (2002). 

\bibitem{SantoriNJP04} C.\ Santori, D.\ Fattal, J.\ Vuckovic, G.\ S.\ Solomon
  and Y.\ Yamamoto, 
New J.\ Phys. {\bf 6}, 89 (July 2004).  

\bibitem{Fasel04} 
S.\ Fasel, O.\ Alibart, S.\ Tanzilli, P.\ Baldi, A.\ Beveratos,
  N.\ Gisin and H.\ Zbinden, 
New J.\ Phys.\ {\bf 6}, 163 (2004).

\bibitem{Pittman04} 
T.\ B.\ Pittman, B.\ C.\ Jacobs, J.\ D.\ Franson,
Opt.\ Comm.\ {\bf 246}, 545-550 (2004).

\bibitem{BerryScheelSandersKnight2004}
D.\ W.\ Berry, S.\ Scheel, B.\ C.\ Sanders, and P.\ L.\ Knight, 
Phys.\ Rev.\ A {\bf 69}, 031806(R), (2004).

\bibitem{BerryScheelSandersKnightLaflamme2004}
D.\ W.\ Berry, S.\ Scheel, C.\ R.\ Myers, B.\ C.\ Sanders, P.\ L.\
Knight and R.\ Laflamme, New J.\ Phys. {\bf 6}, 93 (2004).

\bibitem{Berry05}
D.\ W.\ Berry, A.\ I.\ Lvovsky, B.\ C.\ Sanders, 
preprint: quant-ph/0507216, 2005.

\bibitem{M89} Milburn, G.J., 
  Phys. Rev. Lett. \textbf{62}, (1989).

\bibitem{IHY85}Imoto, N., Haus, H. A., Yamamoto, Y., 
Phys. Rev. A  \textbf{32} 4 (1985).

\bibitem{Munro05} W.\ J.\ Munro, K.\ Nemoto and T.\ P.\ Spiller, 
New J.\ Physics {\bf 7}, 137 (2005).

\bibitem{Konrad05} T. Konrad, A. Scherer, M. Nock and J. Audretsch,
  quant-ph/0510113 (2005).

\bibitem{ScheelKnight2003}
S.\ Scheel, K.\ Nemoto, W.\ J.\ Munro, and P.\ L.\ Knight,
Phys.\ Rev.\ A {\bf 68}, 032310 (2003).

\bibitem{SchmidtImamoglu1996}
H.\ Schmidt and A.\ Imamo$\check{\mbox{g}}$lu, Opt.\ Lett.\ {\bf 21}, 1936, (1996)

\bibitem{LukinImamoglu2000}
M.D.\ Lukin and  A.\ Imamo$\check{\mbox{g}}$lu, Phys.\ Rev.\ Lett.\ {\bf 84}, 1419 (2000).

\bibitem{Harris1997}
S.E.\ Harris, Phys.\ Today {\bf 50}, No.\ 7, 36 (1997).

\bibitem{WangMarzlinSanders2006}
Z.-B.\ Wang, K.-P.\  Marzlin und B.\ C.\ Sanders, arXiv:
quant-ph/0602029 (2006)


\end{thebibliography}
\end{document}